# Resistance control of a magnetoresistive manganite by spin-injection

T. Ono, A. Kogusu, S. Morimoto, and S. Nasu

*Graduate School of Engineering Science, Osaka University, Toyonaka 560-8531, Japan*

A. Masuno, T. Terashima, and M. Takano

*Institute for Chemical Research, Kyoto University, Uji 611-0011, Japan*

**Abstract**

We report a new spin-injection effect found for a manganite using a specially fabricated sample. A wire of $La_{0.7}Sr_{0.3}MnO_3$ was patterned by means of focused ion beam etching, and the central part was subsequently irradiated with $Ga^+$ ions lightly. The ferromagnetic Curie temperature was reduced locally by the irradiation from $T_C$ to $T_C$', and thus a sequential ferromagnetic/paramagnetic/ferromagnetic structure was realized along the wire between $T_C$ and $T_C$'. The injection of spin-polarized current from the ferromagnetic manganite into the paramagnetic part rendered the latter ferromagnetic and more conductive. This can be explained by assuming the suppression of spin fluctuation in the paramagnet by the injected spins.

Spin-injection from a ferromagnetic electrode into a non-magnetic metal or semiconductor is one of the key ideas in spintronics, where new electronic devices based on the spin degree of freedom have been explored[1-4]. Up to now, non-magnetic material in such devices has been simply a medium to transport injected spins from a source to a drain. It is, however, of both fundamental and technological interest if the spin-injection originates drastic changes in physical properties of non-magnetic materials. Here, we report a new spin-injection effect for a magnetoresistive manganite: We show experimentally that injection of spin-polarized current from a ferromagnetic manganite into a paramagnetic manganite renders the paramagnetic counterpart ferromagnetic and more conductive. It is, thus, demonstrated that we can control the resistance by an electric current instead of a magnetic field. Underlying mechanism of this effect is the suppression of spin fluctuation in the paramagnet by the injected spins.

$La_{0.7}Sr_{0.3}MnO_3$ (LSMO) is a well-known conducting ferromagnet[5] of the perovskite type. The coupling of conductivity and ferromagnetism due to a mechanism called double-exchange interaction[6-8] in this oxide leads to two specific properties, which are of great interest from the viewpoint of spin-polarized transport. One is the colossal magnetoresistance (CMR) effect, which is a drastic decrease in resistivity to occur in the presence of external magnetic field suppressing spin fluctuations around the Curie temperature[9]. The other is the half-metallic nature in the ferromagnetic state[10], which makes this and related oxides to be promising candidates for the source of spin-polarized carriers. For the purpose of manifesting spintronic effects using LSMO, we started the present work with the fabrication of a specific sequential structure described below.

A 10 μm-wide wire of LSMO was patterned by means of focused 30-keV $Ga^+$ ion beam (FIB) etching from a 50 nm-thick epitaxial film grown on a $SrTiO_3$ (001) substrate as follows. First, four current-voltage contacts made of silver were prepared on the film by a conventional e-beam lithography and lift-off. The regions around the contacts were then FIB-etched so that a measuring current flows only in the wire. An optical micrograph image of the sample is shown in Fig. 1 (a), where the distance between the two voltage contacts was 10 μm. The central part of the wire was subsequently irradiated with $Ga^+$ ions lightly. This was a way for us to decrease the Curie temperature locally (from $T_C$ to $T_C$') and, thereby, obtain a sequential ferromagnetic/paramagnetic/ferromagnetic (*F/P/F*) structure along the longitudinal axis between $T_C$ and $T_C$'. From this structure we also expected smooth transport of spin-polarized electrons such as illustrated in Fig. 1 (b), because both chemical composition and crystal structure were essentially the same throughout the wire. The $Ga^+$-irradiated area was 0.13 μm × 10 μm in dimension. The ion current was 25 pA and the fluence was $9 \times 10^{16}$ ions/$cm^2$.

Figure 2 shows the temperature dependence of resistance for the irradiated sample and for the non-irradiated sample under various magnetic fields with a measuring current of 0.5 μA ($1 \times 10^2$ A/$cm^2$). The behavior of the non-irradiated sample was similar to that of a bulk crystal: The resistivity was comparable to the reported value[11], the maximum reflecting the onset of ferromagnetism appeared at 340 K, and a CMR effect of 50 % appeared at a magnetic field of 9 T at 340 K. The resistance of the irradiated sample was, however, remarkably larger, and the temperature of the maximum was reduced to 150 K. The CMR effect was not lost, being about 60 % at 4 T at 150 K, but metallicity was lost gradually below 80 K. These results indicated that the irradiation made the Curie temperature as low as $T_C' = 150$ K and the resistivity at the Curie temperature several hundred times as large. It should be noted here that the behavior of this irradiated region is accentuated in the resistance data below ~300 K where the non-irradiated region has a low resistivity. We could thus obtain a desired sequential structure of $F$ (highly conductive, 5 μm in length)/ $P$ (less conductive, 0.13 μm)/ $F$ (highly conductive, 5 μm) along the wire axis between 340 K ($T_C$) and 150 K ($T_C'$).

Intriguing is the contrast between the two samples in the dependence of resistance on the measuring current. Figure 3 shows behaviors of these samples at various measuring currents in the absence of a magnetic field. The current of 1 μA corresponded to a current density of $2 \times 10^2$ A/cm$^2$. The resistance of the irradiated sample showed a strong dependence, while that of the non-irradiated sample did not. The irradiated sample showed the following four features. First, the resistance around $T_C'$ decreases with the measuring current. Second, $T_C'$ rises with increasing measuring current. Third, the current effect vanishes at room temperature. Fourth, a current-induced metal-insulator transition occurs gradually at low temperatures.

The first feature can be explained as follows. Injected into the irradiated region are spin-polarized electrons drawn from source $F$. These electrons of 3$d$ $e_g$ parentage should tend to suppress the fluctuation of the localized 3$d$ $t_{2g}$ spins through intra-atom exchange interactions in region $P$, which, in turn, makes the mobility of the injected electrons higher. This is similar to what happens when an external magnetic field is applied. The increase of $T_C'$ with increasing the measuring current, the second feature, is another expression of the above effect. This effect becomes less remarkable as temperature increases because the injected electrons tend to lose the spin polarization and also because the fluctuation of the localized 3$d$ $t_{2g}$ spins in the irradiated region is enhanced. This is the explanation of the third feature.

To compare the effects of the spin injection and the magnetic field with each other, we plot $T_C$' as functions of the measuring current and the magnetic field in Fig. 4. The $T_C$' was determined as the temperature at which the resistance showed a maximum in Figs. 2 and 3. It is interesting to note that $T_C$' was almost proportional to the measuring current up to 40 µA but gradually saturated at 175 K at high measuring currents. On the other hand, the magnetic field effect was extended to 203 K at 5 T. A scaling relation between the measuring current and the external magnetic field ($5 \times 10^4$ T/A below 40 µA) may be expected from the well-known idea that the effective magnetic field associated with nonequilibrium polarized spins should be in proportion to the injection current[12]. However, this idea cannot explain the fourth feature, *i.e.* the saturation of $T_C$' at high measuring currents. It may be possible that the relaxation time of the injected spins becomes shorter with increasing current density.

An electric field-induced switching from an insulator to a metal is known for a bulk crystal of $Pr_{0.7}Ca_{0.3}MnO_3$[13] to occur at about $10^4$ V/cm, which has been suggested to be a transition from a charge-ordered antiferromagnetic insulator to a ferromagnetic metal. However, we think that this is not the case for the present sample. The irradiated region is artificially defected LSMO, and the defects generate localization potentials which must be random with respect to location, shape, and strength. It is quite natural to assume that the gradual recovery of metallicity comes from the electric field effect overcoming the localization potentials and also from another effect that the magnetic randomness caused by the defects is lightened by the injection of polarized spins.

We thus have shown that the spin-injection into the low-$T_C$ region is like the application of magnetic field. The point of the present work is that this is the first success in the modification of conductivity and magnetism by the use of spin-polarized current. And we emphasize that to this success the fabrication of the specific sequential structure was technically crucial.

We would like to thank S. Maekawa and E. Saito for valuable discussions. The present work was partly supported by the Ministry of Education, Culture, Sports, Science and Technology of Japan (MEXT) through the Grants-in-Aid for Scientific Research (A12304018, A14204070, Scientific Research on Priority Area 407), those for COE Research (10CE2004, 12CE2005) and MEXT Special Coordination Funds for Promoting Science and Technology (Nanospintronics Design and Realization, NDR) and by NEDO project on Nanopatterned magnets.

**Figure captions**

**Figure 1 (a)** Optical micrograph image of the sample. The FIB-etched region and the Ag contacts are seen black and white, respectively. The other part (gray) is $La_{0.7}Sr_{0.3}MnO_3$. **(b)** Schematic illustration of the sample whose central part was irradiated with $Ga^+$. The Curie temperature of the irradiated region is reduced from the as-prepared Curie temperature 340 K ($T_c$) to 150 K ($T_c$') as seen in Fig. 2. The sample has a sequential structure of *F* (ferromagnetic)/*P* (paramagnetic)/*F* (ferromagnetic) along the wire axis between $T_c$ and $T_c$', and spin-polarized electrons are injected from source *F* on the left to region *P*.

**Figure 2** Temperature dependence of resistance for the irradiated sample and the non-irradiated sample under various magnetic fields with a measuring current of 0.5 µA.

**Figure 3** Temperature dependence of resistances for the irradiated sample and the non-irradiated sample at various measuring currents in the absence of a magnetic field.

**Figure 4** Curie temperature ($T_c$') as functions of the measuring current and the magnetic field. The circles and crosses indicate the current dependence and the field dependence, respectively.

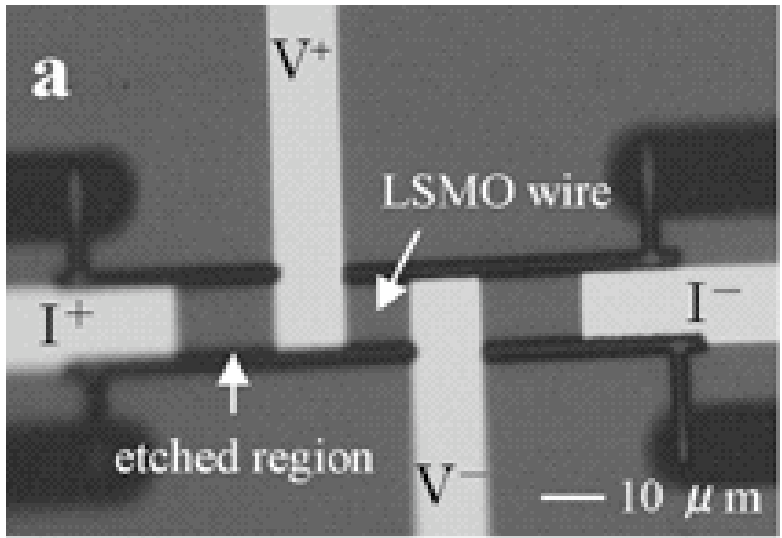

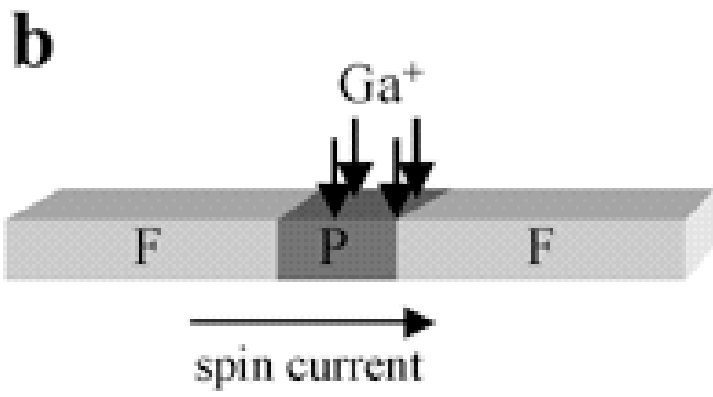

Ono_fig1

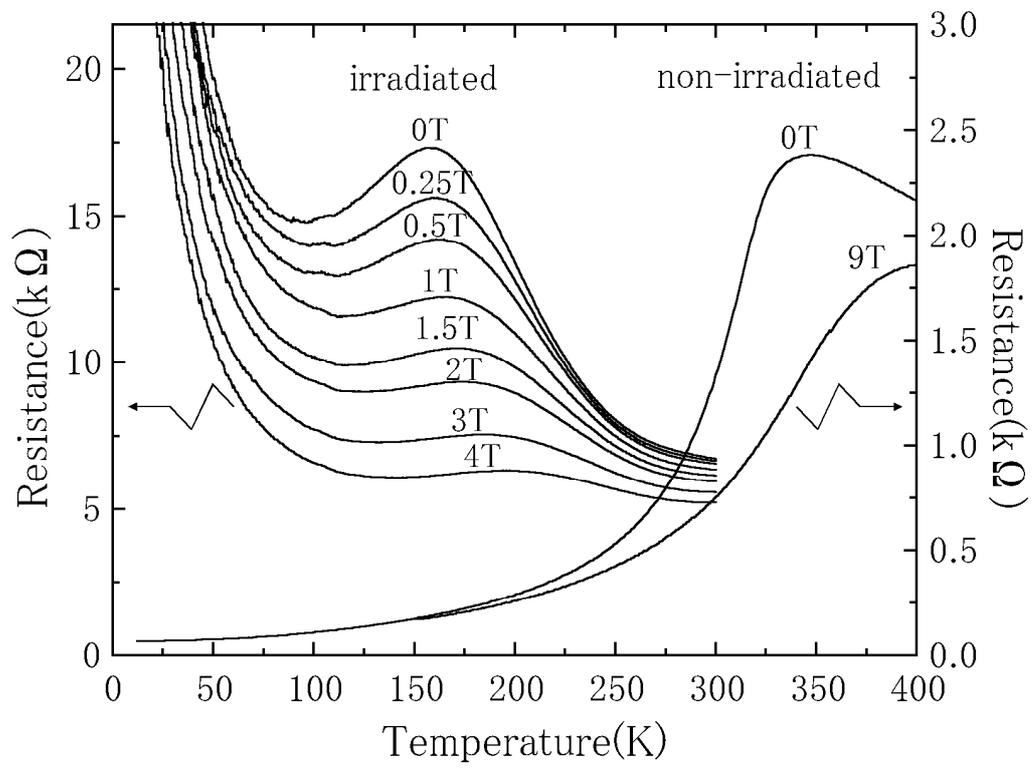

Ono_fig2

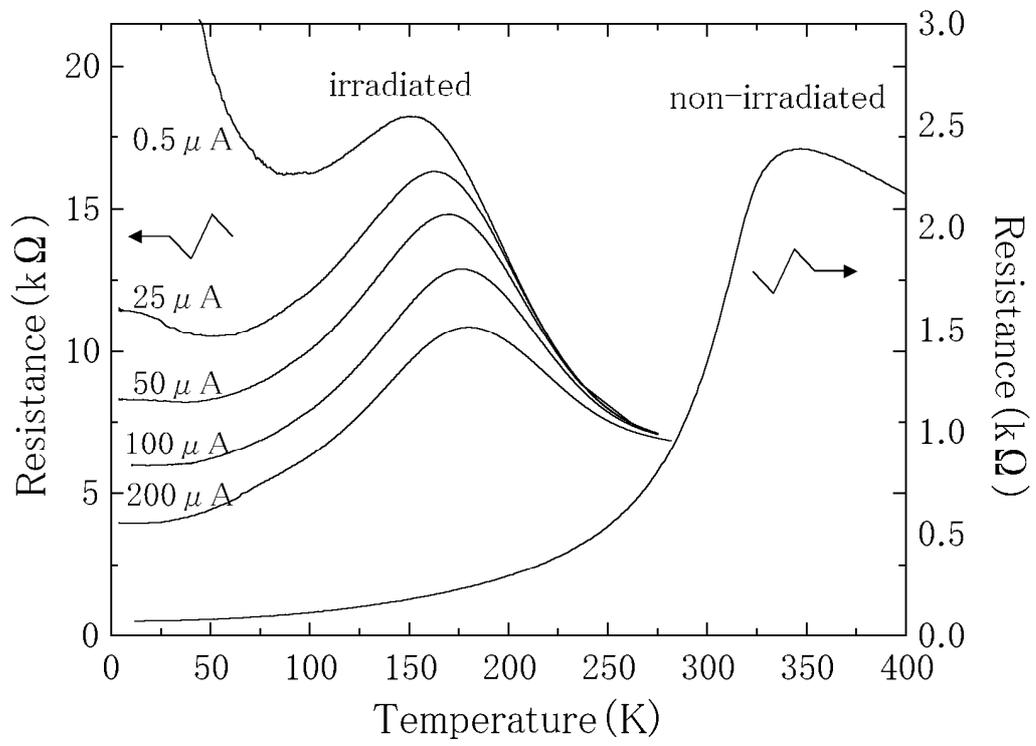

Ono_fig3

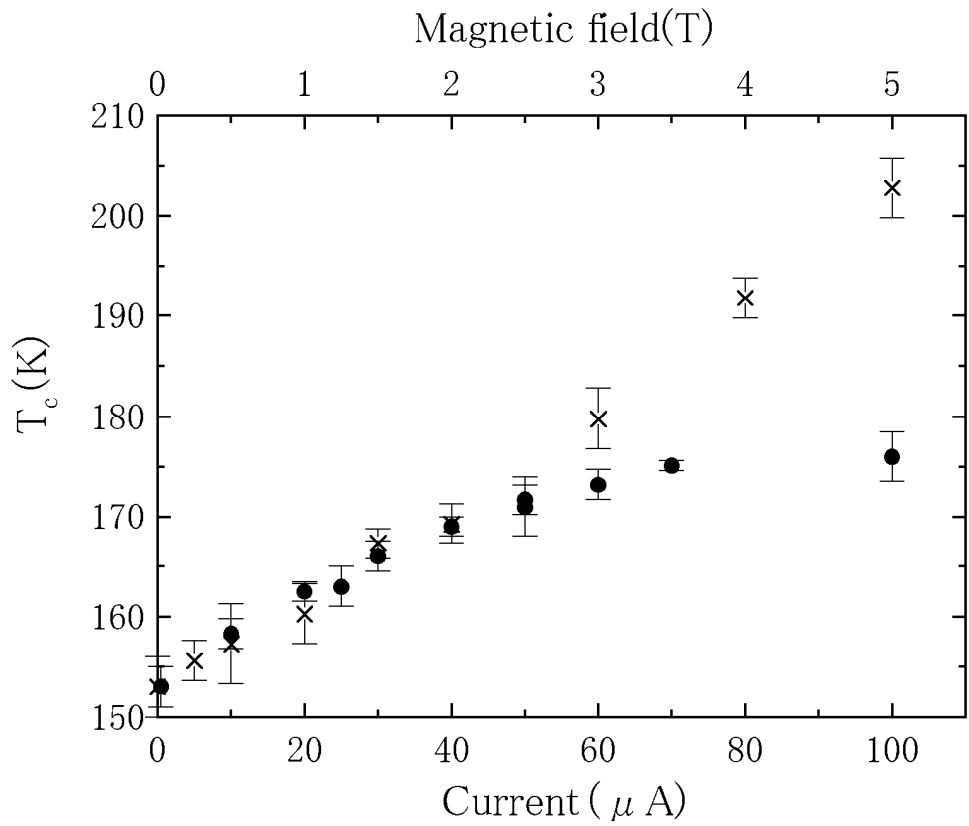

Ono_fig4